\documentclass[traditabstract]{aa}
\usepackage{txfonts}
\usepackage{graphicx}
\usepackage{lscape}
\def\ltsima{$\; \buildrel < \over \sim \;$}
\def\simlt{\lower.5ex\hbox{\ltsima}}
\def\gtsima{$\; \buildrel > \over \sim \;$}
\def\simgt{\lower.5ex\hbox{\gtsima}}

\def\cm2{{cm$^{-2}$}}

\def\p1{{Paper I}}

\def\xmm{{\em XMM--Newton}}
\def\chandra~{{\em Chandra}}

\def\chandra{{\em Chandra}}

\def\xmm{{\em XMM--Newton}}

\def\nh{{N$_{\rm H}$}}

\def\mos{{\em MOS}}

\def\f14{{10$^{-14}$}}
\def\f13{{10$^{-13}$}}
\def\f12{{10$^{-12}$}}
\def\f11{{10$^{-11}$}}
\def\e22{{10$^{22}$}}

\def\3c{{3C 234}}
\def\l58{{$L_{5.8 \mu m}$}}

\begin{document}

\title{HS 1700+6416: the first high redshift non lensed NAL-QSO showing variable high velocity outflows}
\author{G. Lanzuisi\inst{1,}\inst{2} ; M. Giustini\inst{1,}\inst{3}; M. Cappi\inst{1}; M. Dadina\inst{1}; G. Malaguti\inst{1}; C. Vignali\inst{4,}\inst{5}; G. Chartas\inst{6}}
\titlerunning{Relativistic outflows in HS 1700+6416}\authorrunning{G. Lanzuisi et al.}
\offprints{Giorgio Lanzuisi, \email{lanzuisi@iasfbo.inaf.it}}
\institute{INAF - Istituto di Astrofisica Spaziale e Fisica cosmica di Bologna, via Gobetti 101, 40129 Bologna, Italy \and
Max-Planck-Institut f\"ur extraterrestrische Physik,  Giessenbachstrasse 85748 Garching, Germany \and
Center for Space Science and Technology, University of Maryland, Baltimore County, 1000 Hilltop Circle, Baltimore  MD 21155, USA\and
Dipartimento di Astronomia, Università degli Studi di Bologna, via Ranzani 1, 40127 Bologna, Italy \and
INAF - Osservatorio Astronomico di Bologna, via Ranzani 1, 40127 Bologna, Italy \and
Department of Physics and Astronomy, College of Charleston, Charleston, SC 29424, USA}

\date{}

\abstract{We present a detailed analysis of the X-ray emission of HS 1700+6416, a high redshift ($z=2.7348$), 
luminous quasar, classified as a Narrow Absorption Line (NAL) quasar on the basis of its SDSS spectrum.
The source has been observed 9 times by \chandra\ and once by \xmm\ from 2000 to 2007.
Long term variability is clearly detected, between the observations, in the 2-10 keV flux 
varying by a factor of three ( $\sim 3-9 \times 10^{-14}$ erg s$^{-1}$ cm$^{-2}$ ) 
and in the amount of neutral absorption (N$_{H} < 10^{22}$ cm$^{-2}$ in 2000 and 2002 and N$_{H}=4.4\pm1.2 \times 10^{22}$ cm$^{-2}$ in 2007).
Most interestingly, one broad absorption feature is clearly detected at $10.3\pm0.7$ keV (rest frame) in the 2000 \chandra\ observation, 
while two similar features, at $8.9\pm0.4$ and at $12.5\pm0.7$ keV, are visible when the 8 contiguous \chandra\ observations of 2007 
are stacked together. In the \xmm\ observation of 2002, strongly affected by background flares, there is a hint for a similar feature at $8.0\pm0.3$ keV.
We interpreted these features as absorption lines from a high velocity, highly ionized (i.e. Fe XXV, FeXXVI) outflowing gas.
In this scenario, the outflow velocities inferred are in the range v$=0.12-0.59$c.
To reproduce the observed features, the gas must have  high column density (N$_{H}>3\times10^{23}$ cm$^{-2}$),
high ionization parameter (log$\xi>3.3$ erg cm s$^{-1}$) and a large range of velocities ($\Delta V \sim 10^4$ km s$^{-1}$).
This Absorption Line QSO is the fourth high-z quasar displaying X-ray signatures of variable, high velocity outflows, and among these,
is the only one non-lensed.
A rough estimate of the minimum kinetic energy carried by the wind of up to $18\%$ L$_{bol}$,  based on a biconical geometry of the wind, 
implies that the amount of energy injected in the outflow environment is large enough to produce effective mechanical feedback.
}

\maketitle

\section{Introduction}

It is now widely accepted that most nearby bulge-dominated galaxies
may contain a 'relic'  Super Massive Black Holes (SMBHs).
The existence of a tight relation between the mass of the SMBH and the properties 
(mass, luminosity, velocity dispersion) of the host spheroid 
(Kormendy \& Richstone 1995, Silk \& Rees 1998, Magorrian et al. 1998, Ferrarese \& Merritt 2000), 
suggest that:
1) most bulge galaxies went through a phase of strong nuclear activity during 
their formation process and 
2) the evolution of the galaxy and of the SMBH must be closely related, 
implying some sort of interaction or feedback.
Models have shown that luminous AGN might sterilize their host galaxy, 
by heating the interstellar matter through winds, shocks and ionizing radiation, 
strongly inhibiting star-formation in these galaxies and making their colors redder 
(Hopkins et al. 2008, Cattaneo et al. 2009). 
Some observational evidence for such a scenario is starting to be found in a 
handful of objects at high redshift (e.g. Cano-Diaz et al. 2012; Page et al. 2012).

One effective way of providing mechanical feedback can be through massive, wide-angle AGN winds or outflows,
that can be observed as absorption features in optical/UV/X-ray spectra of AGN. 
In particular, UV broad absorption lines (BAL, FWHM$>2000$ km s$^{-1}$, Turnshek et al. 1980) of ionized metals are observed in
10-15\% of optically selected QSOs. Mini-BAL ($500<$FWHM$<2000$ km s$^{-1}$, Hall et al. 2002) and narrow absorption lines 
(NAL, FWHM$<500$ km s$^{-1}$, Weymann et al. 1979; Ganguly et al. 2001) QSO
are observed in up to 30\% of optically selected QSOs.
These absorption features are often strongly blue-shifted with respect to the source redshift, indicating outflowing velocities 
between 10$^3$ and few $\times10^4$ km s$^{-1}$ (Ganguly \& Brotherton 2008) .

In the X-ray band, absorption due to ionized species such as N VI-VII, O VII-VIII,
Mg XI-XII, Al XII-XIII, Si XIII-XVI, as well as L-shell transitions of Fe XVII-XXIV, is
observed to be blueshifted by a few $10^{2}-10^{3}$ km s$^{-1}$ in $\sim50\%$ of type 1 AGN (the "warm
absorber", e.g. Reynolds 1997; Piconcelli et al. 2005; McKernan et al. 2007). 
In the last decade, thanks to the high collecting area of \xmm, \chandra\ and {\it Suzaku}, 
blue-shifted absorption lines due to highly ionized gas (i.e. Fe XXV, Fe XXVI) 
outflowing at high velocity (v$_{out}=1.5-6\times10^4$  km s$^{-1}$) have been observed in a
number of local AGN (Markowitz et al. 2006; Braito et al. 2007; Cappi et al. 2009; Reeves et al. 2009; Tombesi et al. 2010; Giustini et al. 2011). 

\begin{table*}[t]
\begin{center}
\caption{Observation Log. (1) Telescope; (2) Detector; (3) Observation ID; 
(4) Observation date; (5) Exposure in ks (net exposure); (6) Net source counts
in the 0.5-8 keV band.}
\begin{tabular}{cccccc}
\hline\hline\\
\multicolumn{1}{c} {Instrument} & 
\multicolumn{1}{c} {Detector}  &        
\multicolumn{1}{c} {ObsID} &        
\multicolumn{1}{c} {Date}&       
\multicolumn{1}{c} {Exposure (Net)} &     
\multicolumn{1}{c} {Net Counts}  \\
 (1) & (2) &(3) & (4) & (5) & (6)\\  
\hline\\  
\chandra   &   ACIS-I   &         547    &   2000-10-31 &  50.1 (49.5)  &   380   \\  
\xmm       &   PN       &    0107860301  &   2002-05-31 &  27.1 (10.8)  &   312   \\  
\xmm       &   MOS1     &    0107860301  &   2002-05-31 &  27.1 (12.9)  &   145      \\
\xmm       &   MOS2     &    0107860301  &   2002-05-31 &  27.1 (13.2)  &   155      \\
\chandra   &   ACIS-I   &        8032    &   2007-11-12 &  31.4 (31.0) &   205   \\  
\chandra   &   ACIS-I   &        9757    &   2007-11-13 &  21.0 (20.7) &   124   \\  
\chandra   &   ACIS-I   &        9756    &   2007-11-14 &  32.6 (32.3) &   221   \\  
\chandra   &   ACIS-I   &        9758    &   2007-11-16 &  23.6 (23.4) &   194   \\  
\chandra   &   ACIS-I   &        9759    &   2007-11-17 &  31.5 (31.2) &   201   \\  
\chandra   &   ACIS-I   &        9760    &   2007-11-19 &  17.1 (16.9) &   153   \\  
\chandra   &   ACIS-I   &        8033    &   2007-11-20 &  30.0 (29.7) &   204   \\  
\chandra   &   ACIS-I   &        9767    &   2007-11-21 &   9.1 (9.0) &   60    \\  
\hline      
\hline                              
\end{tabular}
\end{center} 
\end{table*}

Finally, X-ray BALs, blueshifted up to v$_{out} \sim2\times10^5$ km s$^{-1}$ , have been observed in gravitationally lensed 
BAL QSOs at high redshifts (Chartas et al. 2002, but see also Hasinger et al. 2002, Chartas et al. 2003, 2009, Saez \& Chartas 2011). 
\begin{figure}[h]
\begin{center}
\includegraphics[width=7cm,height=7cm]{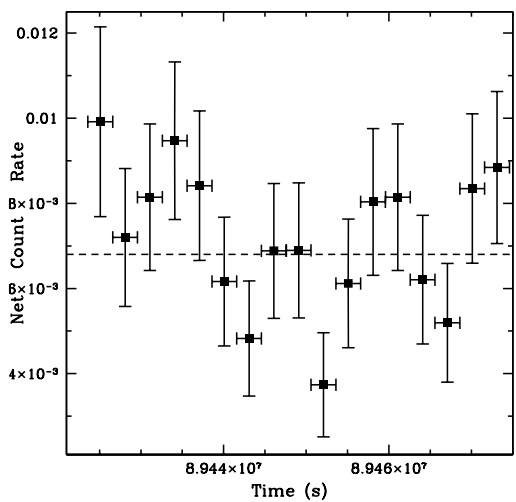}
\caption{\chandra~0.5-8 keV, background subtracted, light-curve of HS 1700+6416 extracted from the 50 ks observation of 2000. The bin size is 3000 s.
The dashed line shows the mean count rate value of $6.8\times10^{-3}$ counts/s.}
\end{center}
\label{lc}
\end{figure}
These lines are associated with a zone of circum-nuclear gas, photo-ionized by the central X-ray source, with 
ionization parameter \footnote{$\xi=L/nR^2$, where L is the ionizig luminosity, n the particle density and R the distance between the 
ionizing source and the gas (Tarter et al. 1969).}
log$\xi\sim3-5$ erg cm s$^{-1}$ and column density \nh$\sim10^{22}-10^{24}$ cm$^{-2}$.
These findings suggest the presence of massive clouds of highly ionized absorbing material,
outflowing at nearly relativistic velocities from the nuclear regions, possibly 
connected with accretion disk winds and/or the base of a jet (e.g. Ghisellini et al. 2004).
However, these features are difficult to observe, 
because they are intrinsically variable, 
the gas is highly ionized (and thus only a few transitions are observable), 
and the spectral range where they occur lies at the high energy edge of the observing 
band available for present high throughput X-ray observatories
(Cappi et al. 2006).

To understand the physical mechanism responsible for 
launching and accelerating AGN winds, it is crucial to build up a large sample
of sources for which these X-ray features can be studied in detail, especially at high redshift,
where the average SMBH accretion rate is higher (Barger et al. 2001).
This will allow for the deeper understanding of the link 
between the accretion and ejection processes, quantifying the kinetic energy injected in the environment, 
and thus the role of outflows in the formation of cosmic structures.
In particular, good quality spectra of high-z QSO showing broad absorption features in the iron K band,
are limited to three gravitationally lensed BAL and mini-BAL QSOs (APM
08279+5255, PG 1115+080 and  H 1413+117, Chartas et al. 2002, 2003, 2007).
These sources are the only few optical/UV BAL quasars to date that are
bright enough in the X-ray band, mostly due to the strong lensing magnifications, 
to allow in-depth spectral analysis.

Here we present a detailed analysis of the X-ray emission of the quasar HS 1700+6416, a high redshift (z=2.7348), 
high luminosity NAL QSO.
HS 1700+6416 is one of the most luminous, non-lensed (Reimers et al. 1997), QSOs in the SDSS ($M_i=-30.24$).
It is not detected in the NVSS radio survey, that has a 5 $\sigma$ sensitivity limit of 2.5 mJy at 1.4 GHz.
The source should be therefore considered extremely radio quiet (log(F$_{1.4GHz}$ /F$_R$ $< -5$)\footnote{F$_{1.4GHz}$ is the 1.4 GHz radio flux density})).
HS 1700+6416 is classified as a NAL QSO, showing narrow CII, CIV, SiIII and Si IV absorption lines in the SDSS spectrum,
blue shifted at mildly ($\sim2.4\times10^4$ km s$^{-1}$) relativistic velocities (Misawa et al. 2007). 
The source has a very faint ROSAT counterpart (Reimers et al. 1995).

The paper is organized as follows: 
Section 2 reports on the X--ray observations and data reduction; Section 3 presents the spectral analysis, while
the main results are discussed in Section 4.
A standard $\Lambda$ cold dark matter cosmology with $H_0=70$ km s$^{-1}$ Mpc$^{-1}$, $\Omega_\Lambda=0.73$ and $\Omega_M=0.27$ is assumed throughout the paper.
Errors are given at 90\% confidence level ($\Delta C=2.706$ for one parameter of interest), unless otherwise specified.

\section{X-ray data analysis}

\subsection{X-ray observations}

The quasar HS 1700+6416 lies in a particularly dense and interesting field in the sky.
It is close to two clusters, Abell 2246, at $z=0.225$ and 2.2' of angular separation, and V1701+6414, at $z=0.45$ and 3.2' of angular separation
(Vikhlinin et al. 1998), and has a foreground over-density of galaxies and AGN, centered at $z=2.300 \pm 0.015$ (Digby-North et al 2010).
For this reason, the field has been observed by \chandra\ and \xmm\ several times (Tab. 1), with different scientific goals.
In particular, one 50 ks \chandra\ observation was performed in 2000, to study the cluster V1701+6414 (P.I.: L. Van Speybroeck);
one 27 ks \xmm~observation in 2002 was performed to study the same cluster (P.I.: F. Jansen), and 
a series of 8 \chandra\ observations,  for a total of $\sim200$ ks, 
was taken during 9 days in 2007, to study the rate of AGN activity 
among the galaxies in the over-density at z=2.300 (P.I.: K. Nandra).

The 50 ks \chandra\ spectrum of HS 1700+6416 was presented in Just et al. 2007, in a systematic study 
of the X-ray properties of the most luminous quasars in the SDSS.
The authors applied the Cash statistic to the source unbinned data, 
and presented the spectra binned at a level of 10 counts per bin, for clearer presentation. 
In the spectrum of HS 1700+6416 they found two consecutive bins falling $\simgt3\sigma$ below a simple power-law model, around 3 keV (11 keV 
rest frame). They claim the possibility of this being an absorption feature, and reported an improvement of the fit, with the addition of 
an absorption edge at 2.4 keV (9.0 keV in rest frame), significant at a confidence level $>99.6\%$. 

Misawa et al. 2008 presented an exploratory study of the X-ray properties of NAL-QSOs, analyzing the 50 ks \chandra\ spectrum and the 
\xmm\ spectrum. However they binned the \chandra\ spectrum to 20 counts per bin and the \xmm\ pn and MOS to 150 and 45 counts per bin, respectively,
at E$>2$ keV. This led to the loss of any information about the presence of absorption features in the spectra.
Here we present the re-analysis of these two data sets, with additional 8 \chandra\ observations, 
with exposure times in the range 9-32 ks.

\begin{figure}[t]
\begin{center}
\includegraphics[width=6cm,height=6cm]{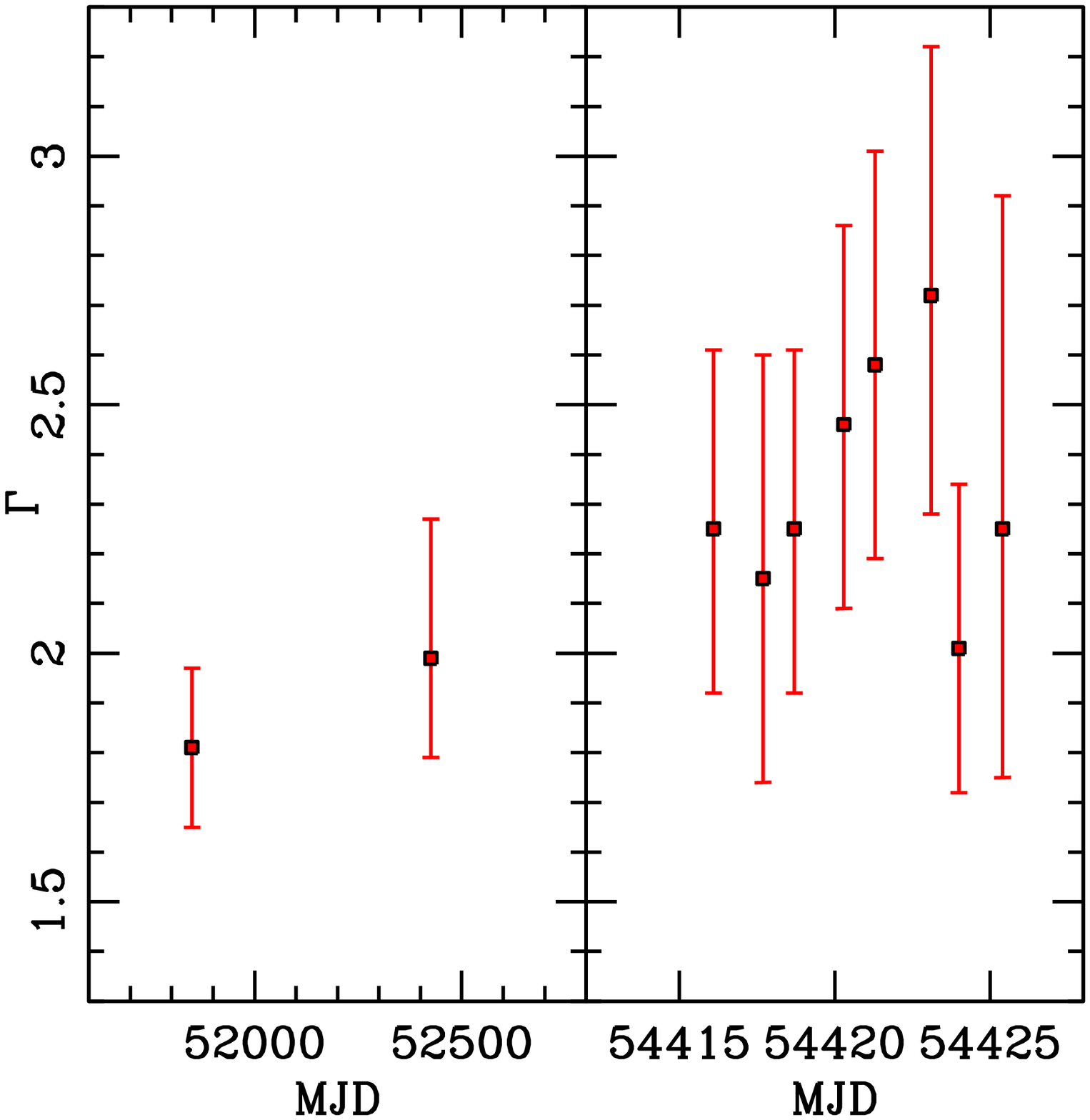}\hspace{1.cm}
\includegraphics[width=6cm,height=6cm]{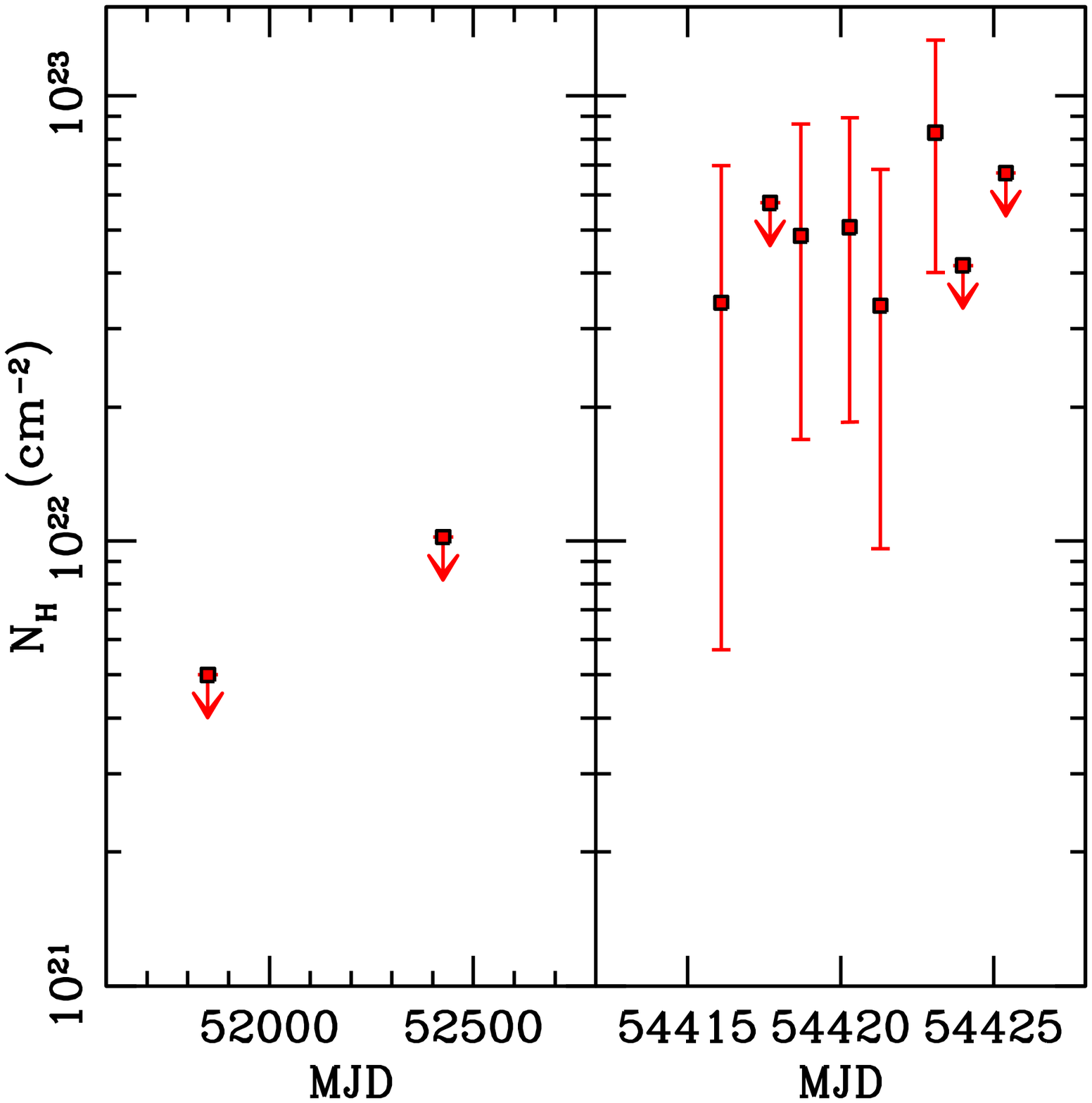}\hspace{1.cm}
\includegraphics[width=6cm,height=6cm]{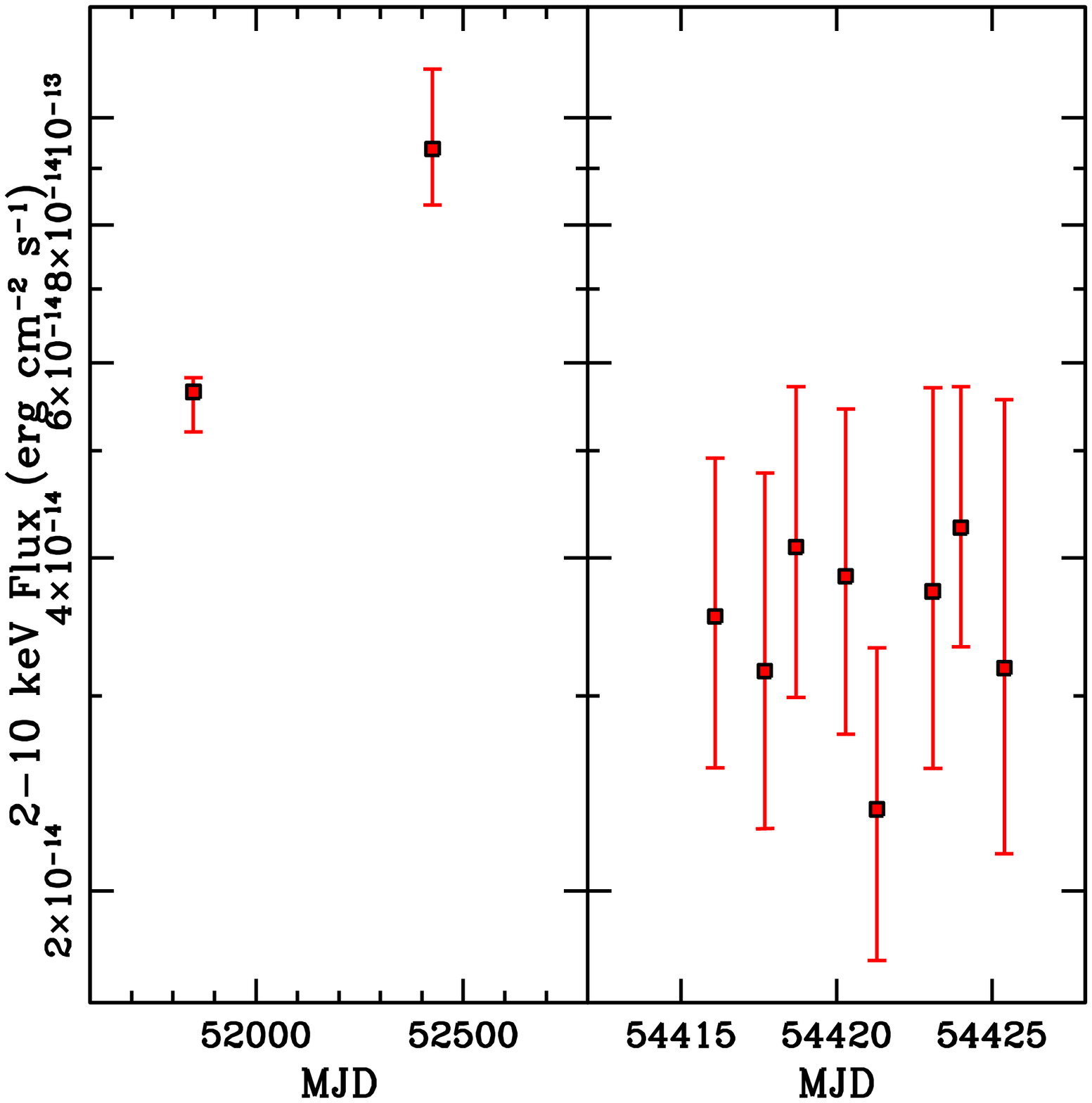}
\caption{From top to bottom panel: Photon index $\Gamma$, column density \nh and 2-10 keV flux, in different observations. Time is in Modified Julian Date.
The x axis scale of the right part of each panel is expanded to make visible the 8 contiguous observations of 2007.}
\end{center}
\label{var}
\end{figure}

\subsection{X-ray data reduction}

The event files of the \chandra\ observations were retrieved from the \chandra\ X-ray Center (CXC) via the Web ChaSeR 
(Chandra Archive Search and Retrieval) interface.
The data reduction was performed following the standard procedures outlined in the Science Analysis Threads for ACIS data at the CIAO (v 4.3) Web 
site\footnote{\textit{http://cxc.harvard.edu/ciao}}.
The 8 repeated observations of 2007 were meant to study the rate of AGN activity in the proto-cluster at z=2.3, and the quasar HS 1700+6416 was observed only as a
by-product. As a result the source falls close to the ACIS-I ccd gap region, in all the 8 observations (with identical aim point).
As a consequence, the effective total exposure at the source position is reduced by $\sim20\%$ due to dithering.

The \xmm\ data were retrieved from the XMM-Newton archive\footnote{\textit{http://xmm.esac.esa.int}}. 
Standard \xmm\ SAS tasks \textit{epproc} and \textit{emproc} (SAS v. 7.1.0)\footnote{\textit{http://xmm.esac.esa.int/sas/}} 
were used to produce calibrated pn and \mos~event files.
The event files were processed using the latest calibration files
and cleaned up from effects of hot pixels and  cosmic rays contamination.
X-ray events corresponding to patterns 0--12 (0--4) for  the \mos (pn) cameras were selected.
The \xmm\ observation was almost completely affected by background flares 
(see the \xmm\ User Handbook).
We produced a global light-curve at energies $>$10 keV, where the contribution from the emission of real X-ray sources is minimized.
Different count rate cuts were performed, in order to find the best trade-off between increasing the number of source counts available and 
decreasing the signal to noise ratio.
The final cut produced cleaned exposures of $\sim10$ and $\sim13$ ks for pn and MOS, respectively.

We extracted the source counts from circular regions of radius 4\arcsec for the \chandra~pointings.
We tried different extraction regions for the \xmm~EPIC cameras, in order to minimize the background contribution: 
15\arcsec, 25\arcsec and 40\arcsec, i.e. 70, 80 and 90\% of the Encircled Energy Fraction (EEF) respectively. 
Given the relative flux level of source and background,  
we chose the smallest extraction region (15\arcsec), that minimizes the contribution of the background,
that could otherwise strongly affect the spectrum, especially above 2 keV.

The background counts were extracted from annular regions close to the source, avoiding ccd gaps and other sources.
The extraction areas were typically $\sim10$ times the source extraction region, in order to average out any position-dependent background features.
The obtained spectra have a number of source net counts, in the 0.5-8 keV band, of $\sim400$ for the 50 ks \chandra~observations,
$\sim310 (300)$ net counts for the pn (MOS1 and MOS2 combined), and in the range 80-220 counts in each of the 8 repeated \chandra~observations.

Given the exposure times of the different observations, and the flux level of the source, the study of the short term variability is
feasible only for the long \chandra\ exposure performed in 2000.
Fig. 1 shows the 0.5-8 keV background subtracted light-curve of HS 1700+6416, 
obtained from the 50 ks \chandra\ exposure, with a bin size of 3000 s, result in $>20$ net source counts in each bin.
When fitted with a constant, the resulting count rate is $6.8\times10^{-3}$ counts s$^{-1}$, with $\chi^2/d.o.f.$ = 21.4/16, and
null hypothesis probability P($\chi^2/\nu$)= 0.16. 
Thus, the source is found consistent with being constant on time scales shorter than the duration of the observation (50 ks).

\section{Spectral analysis}

In the following, we will describe in detail the spectral analysis performed for the source:
first, we determine the properties of the continuum, with a simple absorbed power law model; then 
we focus on the parameters of the strong absorption feature detected in at least 2 of the 4 available spectra, 
using a phenomenological model (a Gaussian absorption line superimposed to the continuum);
finally (Sec. 4) we investigate the possibility of the feature being produced by an absorption edge or by a highly ionized 
outflowing gas.

\begin{figure*}[t]
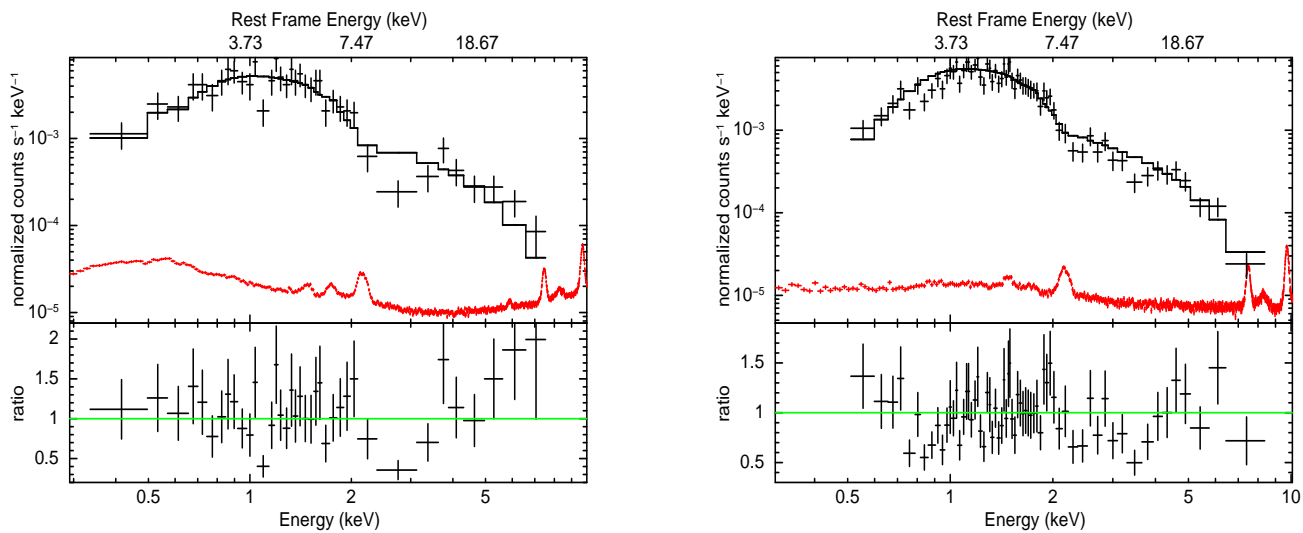

\begin{center}
\includegraphics[width=7cm,height=9.1cm,angle=-90]{547_source_back.ps}\hspace{0.1cm}
\includegraphics[width=7cm,height=9.1cm,angle=-90]{merged_source_back.ps}
\caption{{\it Left panel:} Observed \chandra\ 50 ks spectrum and residuals of HS 1700+6416, modeled with a power law modified by Galactic absorption plus
intrinsic neutral absorption. The counts are binned to a minimum significant detection of $3\sigma$ (and a maximum of 15 counts per bin), only for plotting purposes.
{\it Right panel:} Observed \chandra\ spectrum and residuals of the source obtained from the 8 observations of 2007.
The same model and binning strategy of left panel are used. In red is shown the \chandra\ background.}
\end{center}
\label{cha_spec}
\end{figure*}

Given the low number of source counts available in each observation, and in order to investigate the presence and properties of 
a possible absorption feature in the X-ray spectra of the NAL QSO HS 1700+6416,
we performed all the spectral fits using the Cash statistic (Cash 1979) implemented in Xspec v12.
This is a maximum likelihood function, that does not require counts binning, allowing the full exploitation of
the \chandra\ and \xmm\ cameras' energy resolution.
The Cash statistic assumes that the error on the counts is purely Poissonian, and cannot therefore deal with background subtracted data.
For this reason, a careful characterization of the background spectrum is needed, 
before performing a global fit to the source+background spectrum.

The \chandra\ global background between 0.5 and 7 keV is complex,
and has been reproduced with two power laws, modified by photoelectric absorption, plus 3 narrow Gaussian lines 
to reproduce the features at 1.48, 1.74 and 2.16 keV (Al, Si and Au instrumental lines), plus a thermal (MEKAL) component, 
to account for the sum of the particle, cosmic and galactic components of the background (Markevitch et al. 2003, Fiore et al 2012).
The \xmm\ EPIC pn (MOS) background between 0.5 and 10 keV, has been reproduced with a similar model, 
but with 2 narrow Gaussian lines to account for the prominent features
at 1.5 and 8 (1.8) keV (Nevalainen, Markevitch \& Lumb 2005).
In each observation the background shape has been recovered from background spectra extracted from a large (8 arc-min) region, excluding the astrophysical sources,
in order to collect a large number of counts, thus allowing a detailed, global characterization of the background.
Then, each model has been rescaled and fitted to the local background, leaving as free parameters the overall normalization and the photoelectric absorption, 
to account for position dependent variations.

\subsection{Continuum and Variability}

The source spectra have been fitted with a simple power-law, modified by intrinsic neutral absorption at the source redshift, 
plus galactic absorption (model {\it wabs*zwabs*po} in Xspec), 
fixed to the value measured by Kalberla et al. (2005) at the source coordinates \footnote{http://heasarc.nasa.gov/cgi-bin/Tools/w3nh/w3nh.pl} 
($N_{H,Gal.} = 2.3\times10^{20} cm^{-2}$). 

Fig. 2 shows the distribution of spectral parameters $\Gamma$\footnote{defined as $F\propto E^{-\Gamma}$, the power law spectral index.}, \nh
\footnote{defined as $M(E)=exp(-N_h \sigma(E))$, the equivalent hydrogen column density of the absorber, where $\sigma(E)$ is the photo-electric cross-section.}
and the 2-10 keV flux, as a function of time (MJD), for all the spectra available.
The x axis scale of the right panel of each plot is expanded, with respect to the left panel, to make visible the results obtained for each of 
the 8 contiguous observations in 2007.

Long term 2-10 keV flux variability by a factor $\sim3$ is clearly detected, from $\sim9\times10^{-14}$ erg s$^{-1}$ cm$^{-2}$ in 2002, to 
$\sim3.5\times10^{-14}$ erg s$^{-1}$ cm$^{-2}$ in 2007.
Also the amount of neutral absorption varied, from \nh$<10^{22}$ cm$^{-2}$ in 2000 and 2002, 
to $4-8\times10^{22}$ cm$^{-2}$ in 2007.
The photon index is consistent with being constant, within the large error bars.

The time span between the 8 observations performed in 2007 is 9 days.
In this period the spectral parameters of the source appear to be constant, within the error-bars,
allowing us to stack together all the spectra in order to obtain an average, better quality spectrum. 
We summed all the source and background spectra, and produced weighted ARF and RMF with 
{\it mathpha}, {\it addarf} and {\it addrmf} tools, respectively, 
included in the HEASoft suite\footnote{http://heasarc.nasa.gov/docs/software/lheasoft}.
The matrices are weighted on the basis of the exposure time, however we stress the fact that the 
8 observations have exactly the same aim point, and thus the different ARFs and RMFs are almost equivalent.
As a further check, we also stacked together the 8 event files, and extracted a single set of spectra and matrices, 
obtaining consistent results.

HS 1700+6416 also has a rough mass estimate of M$_{BH}$=$2.5\times10^{10}$ $M_{\odot}$ (Shen et al. 2011). 
We stress that this extreme value, obtained from the C IV emission line,
has large uncertainties, 
due to the possible presence of a non virialized (i.e. a disk wind) component in the CIV line (Shen et al 2008, 2011; Ho et al. 2012),
and should be considered as an upper limit.
We can roughly estimate the BH mass also from the 2-10 keV luminosity.
The 2-10 keV intrinsic luminosity of the source goes from $7.27\times10^{45}$ erg s$^{-1}$ of 2002
to $3.85\times10^{45}$ erg s$^{-1}$ of 2007.
From the highest value, assuming that the BH is accreting at the Eddington limit, and using a 
bolometric correction factor of 50 (Vasudevan \& Fabian 2007), 
it is possible to derive a lower limit for the
BH mass of $3.9\times10^{9}$ $M_{\odot}$ that, together with the value reported in Shen et al. (2011) 
gives an indication of the possible BH mass range for the source.
Considering the highest available mass estimate,
we can derive a Schwarzschild radius (r$_{S}=2GM_{BH}/c^2$) of $\sim7.5\times10^{15}cm$ and a light crossing time of 
$\sim250$ ks in the source frame, which correspond to a variability time scale 
of $\sim10$ days in the observer frame, while for the lower value of the BH mass,
the observed variability time scale is of 1.6 days.

\subsection{Absorption features}

Fig. 3, left panel, shows the spectrum of HS 1700+6416 obtained from the 50 ks \chandra\ observation. 
For plotting purposes only, the spectral counts are binned so as to have a minimum of significant detection of $3\sigma$,
or 15 channel per bin otherwise.
The fit to a simple absorbed power-law model shows significant residuals around $\sim2.7$ keV (observed frame)
suggesting the presence of a strong absorption feature.
The C-stat value for this simple model is 137.3 for 170 d.o.f.
When adding an absorption Gaussian line, with energy, line width and normalization as free parameters, 
the resulting C-stat is 117.0  for  167 d.o.f, i.e. a $\Delta C$ of 20.3 for 3 additional parameters 
(we remember that the change in C-stat from one fit model to the next ($\Delta C$) is distributed approximately as $\Delta\chi^2$
in this range of counts).

Performing an F-test we find that the addition of an absorption Gaussian line improved the fit at a confidence level
$>99.999\%$.
To further assess the statistical significance of this result, 
extensive (10000 trials)  Monte-Carlo simulations have been carried out, producing simulated spectra with the FAKEIT
routine in Xspec.
The input model is an absorbed power-law with the best fit parameters obtained from the observed data.
The simulated spectra are fitted at first with the simple power-law model, and then adding an absorption Gaussian line,
with energy, width and intensity free to vary (range of $E_{line}=0.5-7$ keV).
None of the simulated spectra shows a $\Delta C=20$ (the higher value being $\sim15$) and therefore we estimate
that the probability to detect by chance the observed  $\Delta C$ is $<1\times10^{-5}$ 
(i.e. the feature is significant at confidence level $>99.999\%$, i.e. $>4\sigma$), confirming the results obtained with the F-test.
We stress the fact that, if the $\chi^2$ statistic and count binning (20 counts per bin) are applied, we obtain the following results:
$chi^2/d.o.f.$=44.3/20 for the absorbed power-law model, and $chi^2/d.o.f.$=17.9/16, when an absorption Gaussian line is added.
This turns into a slightly lower significance of 99.6\%, in agreement with the value found in Just et al. 2007.

As a further check we produced the ratio between the observed spectrum of HS 1700+6416, and the observed spectrum
of a real featureless source, in order to exclude that some instrumental effect could be responsible for the broad feature at 2-3 keV.
We selected the brightest source in the sample of BL Lacs observed in Donato et al. 2003, i.e PKS 2005-489, with a net count rate of 0.49 count/sec,
because it has a featureless and flat spectrum and was observed with the same instruments (ACIS-I) and in the same period of the observation of HS 1700+6416.
Fig. 5 shows the ratio between the two spectra. 
PKS 2005-489 has a featureless, flat spectrum with $\Gamma=2.2$, i.e. softer than the continuum of HS 1700+6416, as can be seen in Fig. 5 from the  non flat ratio.
More important, the  feature between 2 and 3 keV is clearly visible in the  HS 1700+6416/PKS 2005-489 ratio, confirming that this 
is not an instrumental feature.

\begin{figure}
\begin{center}
\includegraphics[width=7cm,height=8cm,angle=-90]{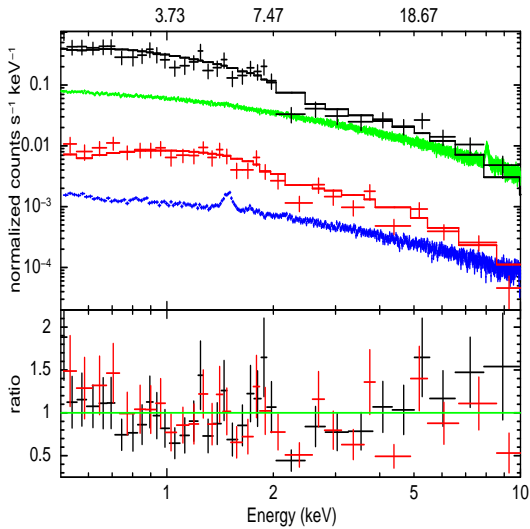}
\caption{Observed \xmm\ pn (black) and MOS1+MOS2 (red) spectra and residuals of the source, modeled with a power law modified by Galactic absorption plus
intrinsic neutral absorption. In green (blue) is shown the pn (MOS) background.}
\end{center}
\label{xmm_spec}
\end{figure} 

\begin{figure}
\begin{center}
\includegraphics[width=6.5cm,height=7cm,angle=-90]{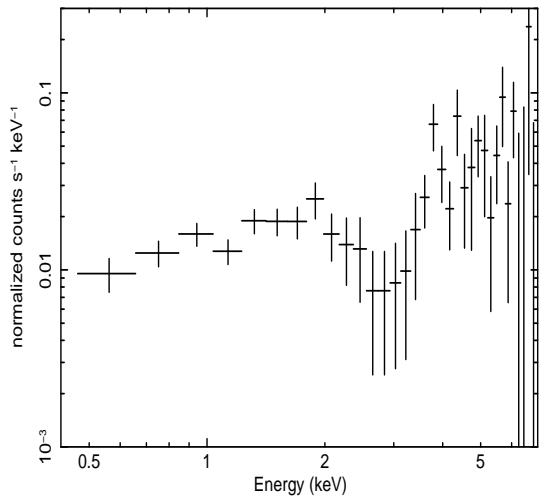}
\caption{Spectrum of the ratio HS 1700+6416 / PKS 2005-489 for the 50ks \chandra\ observation (Cha$_{2000}$ in table 2). the spectra have been binned to $\sim200$ eV bins.}
\end{center}
\label{ratio}
\end{figure} 

The same analysis described above, was performed for the sum of the 8 \chandra\ observations of 2007.
Fig. 3 right panel shows the stacked spectrum.
The counts were binned as for the left panel, for plotting purposes.
The fit to a simple power law plus neutral absorption model, in the 0.5-7 keV band, shows residuals around 2.4 keV, 
and a stronger feature at 3.3 keV (observed frame).
The  C-stat/d.o.f is 219.9/235 for the simple power-law model, while
$\Delta C$ are 4.2 and 16.8 for the 2.4 and 3.3. keV features, respectively, for 3 additional parameters. 
This results in a confidence level of 81.50\% and 99.87\%, respectively, when computed with F-test.

We performed the same type of Monte-Carlo simulations adopted for the long \chandra\ exposure, finding that the 
probability to detect by chance the observed  $\Delta C$ is $1.03\times10^{-1}$ and $6\times10^{-5}$ respectively (significance level 89.72\% and 99.94\%).
Thus, the detection of the feature at higher energies is significant at $>3\sigma$ confidence level, while the 
presence of the feature at lower energies is only marginal ($<2\sigma$).

Finally, Fig. 4 shows the EPIC pn and MOS spectra extracted from the \xmm\ observation of 2002.
The MOS1 and MOS2 spectra and responses were added with {\it mathpha}, {\it addarf} and {\it addrmf} tools.
As already mentioned, the observation was almost completely affected by background flares, and the cleaned exposure time
is of $\sim10$ ks. As a consequence, the number of net source counts available in the 0.5-10 keV band is $\sim300$ for 
each spectrum (pn and combined MOS).

Despite the poor data quality, a hint of presence of an absorption feature 
around $\sim2.1$ keV can be seen in the residuals of both pn and MOS cameras.
In this case the C-stat/d.o.f is 493.2/529 for the simple power-law model, while
$\Delta C$ is 14.3 for 4 additional parameters (line energy, width and 2 normalizations).
This results in a confidence level of 97.50\% (F-test).
Similar results were obtained with Monte-Carlo simulations (98.15\% confidence level).
Thus the evidence of the absorption feature in the \xmm\ spectra was marginal.

\begin{figure*}[t]
\begin{center}
\includegraphics[width=8cm,height=7cm]{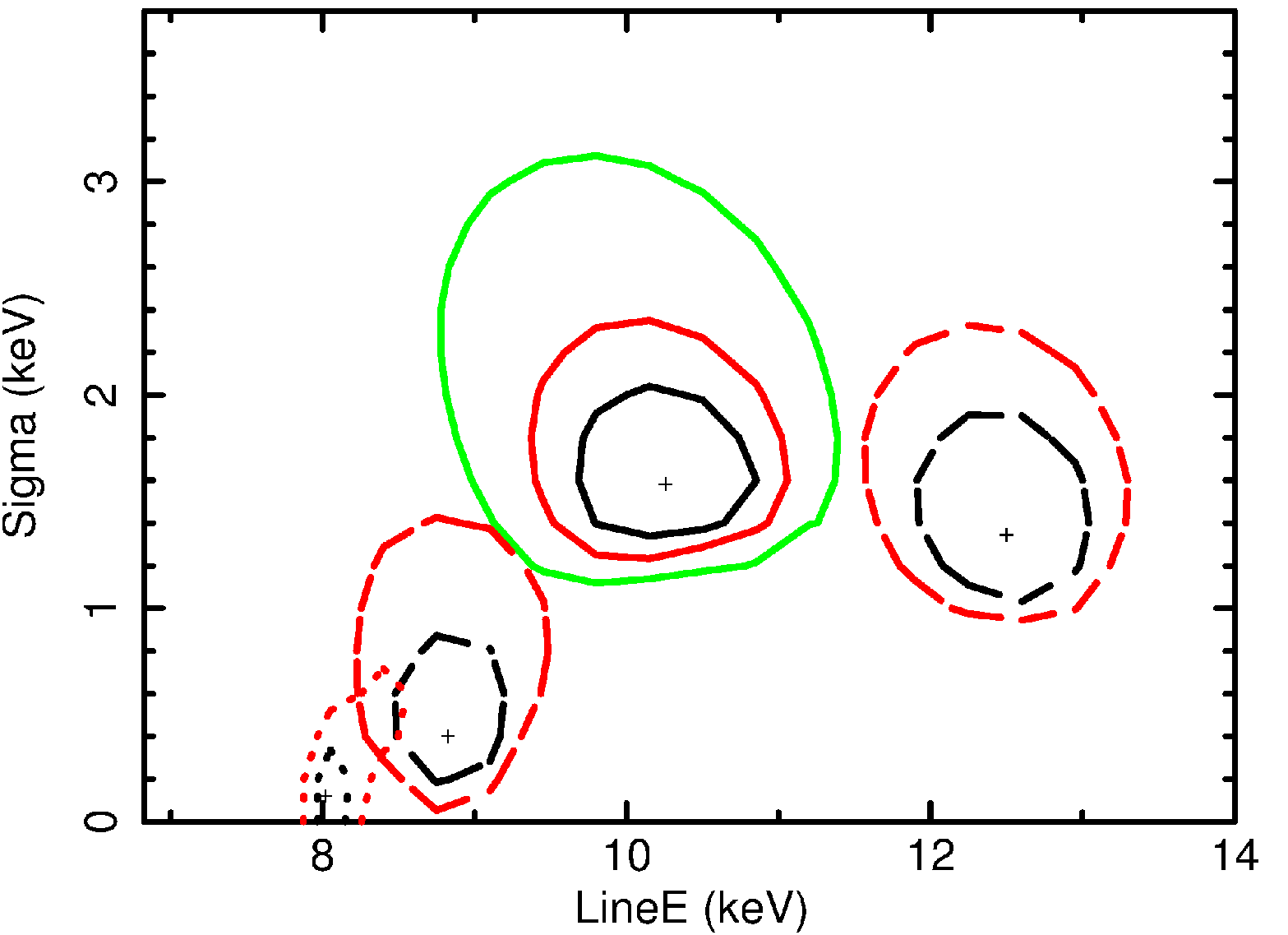}\hspace{1.5cm}
\includegraphics[width=8cm,height=7cm]{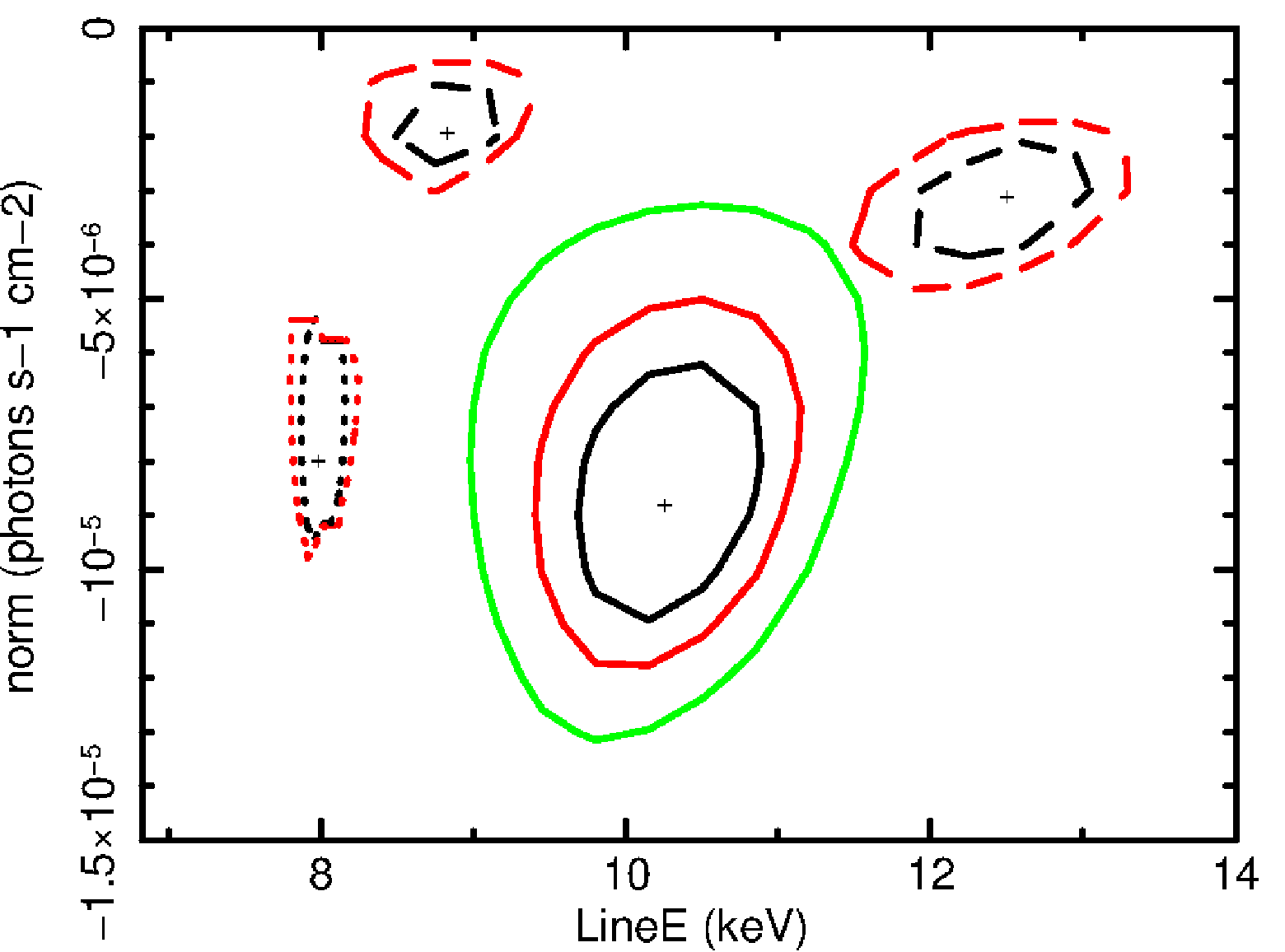}
\caption{{\it Left panel:} 68,90,99\% confidence contours of $E_{line}$ vs. the line width from the APL+Gauss model fit of the \chandra\ 50 ks observation spectrum. 
The dashed (dotted) contours represent 68 and 90\% confidence contours of the parameters of the two lines in the 2007 stacked \chandra\ (2002 \xmm) spectrum.
{\it Right panel:} 68,90,99\% confidence contours of  $E_{line}$ vs. normalization from the APL+Gauss model fit of the \chandra\ 50 ks observation spectrum.
The dashed (dotted) contours represent 68 and 90\% confidence contours of the parameters of the two lines in the 2007 stacked \chandra\ (2002 \xmm) spectrum.}
\end{center}
\label{cont}
\end{figure*}

\section{Modeling of the absorption features}

We tried different models to reproduce the absorption features in the HS 1700+6416 spectra. 
The results are summarized in Tab. 2 for the different models and data sets.
Model 1 is the simple power-law, modified by intrinsic neutral absorption at the source redshift, 
plus galactic absorption, discussed above. 

\subsection{Absorption Gaussian line}

Model 2 is the same model 1 plus one absorption Gaussian line (two for the \chandra\ 2007 spectrum).
Even if this model represent only a phenomenological model to assess the feature significance,
the properties of the absorption lines,
such as the line centroid energy, the width $\sigma$ and the equivalent width (EW hereafter),
can be roughly recovered. 

The rest frame line centroid energy ($E_{line}$) for the 50 ks \chandra\ spectrum is $E_{line}=10.32_{-0.82}^{+0.71}$ keV, 
the line width $\sigma=1.59_{-0.35}^{+0.77}$ keV and the equivalent width $EW=-0.83_{-0.26}^{+0.30}$ keV (all quantities are rest frame). 
Fig. 6 left and right panels, show the 68/90/99\% confidence contours of $E_{line}$ vs. $\sigma$ and $E_{line}$ vs. normalization, respectively, 
for the 50 ks \chandra\ spectrum (solid curves).
Assuming that the absorption feature is due to either Fe XXV $K\alpha$ or Fe XXVI $K\alpha$, 
the two strongest absorption line for an highly ionized gas (Chartas et al. 2009, Saez et al. 2009), 
with rest frame energies of 6.70 keV and 6.97 keV respectively, 
we compute the minimum and maximum outflowing velocity,
taking also into account the $1\sigma$ error on $E_{line}$. 
The resulting velocity is in the range v=0.30-0.46c.

The observed width of the line in this spectrum is extremely large.
This huge broadening can be due to either high turbulence velocity of the absorbing gas, or to the blending of a series 
of absorption lines, corresponding to different gas shells at different ionization states and/or moving at different velocities (or a combination of the three effects).
We underline the fact that the energy of the Fe XXV $K\alpha$ and Fe XXVI $K\alpha$ lines being very close ($\Delta E_{line}$=0.27 keV), 
the simple blending of these two lines cannot be responsible for all of the broadening observed.

The stacked \chandra\ spectrum of 2007 shows a weak feature at $\sim2.4$ keV and a more prominent one around $\sim3.3$ keV.
In the rest-frame, the line parameters are $E_{line1}=8.86_{-0.43}^{+0.39}$ keV and $E_{line2}=12.50_{-0.74}^{+0.64}$ keV, 
$\sigma1<0.95$ keV  and $\sigma2=1.34_{-0.61}^{+0.98}$ keV, 
$EW1=-0.14_{-0.12}^{+0.09}$ keV and $EW2=-0.51_{0.29}^{+0.23}$ keV respectively.
The dashed curves in fig. 6 represent the 68 and 90\% confidence contours of the parameters of the two lines in the 2007 stacked \chandra\ spectrum.
The resulting outflowing velocities are in the range v1= 0.20-0.31c and v2= 0.48-0.59c, respectively.
Finally, the line parameters for the \xmm\ pn and MOS spectra fitted together are $E_{line}=8.05_{-0.22}^{+0.33}$ keV, $\sigma<0.42$ keV and $EW=-0.13_{-0.04}^{+0.07}$ keV.
In this case the resulting outflowing velocity is v= 0.12-0.22c.
A hint of the presence of an emission line, close to the energy of the neutral Fe K$\alpha$ line, is present both in the \chandra\ (2007) and in the \xmm\ spectrum.
However the feature is not statistically significant in either of them, and including such line in the spectral fit did not affect 
the properties of absorption feature.

Fig. 6 clearly shows that the absorption feature is highly variable in energy, intensity and width. 
In particular, if we consider only the two features with the higher significance, i.e. the ones at 10.3 and 12.5 keV respectively, the 
probability that the energy of the first absorption line is the same of the second is $< 1\times10^{-3}$,
i.e. the feature is variable in energy at a confidence level of $> 99.9\%$.

\begin{table*}
\caption{Results from spectral fits.Model APL is a power law with Galactic absorption and intrinsic absorption (XSPEC model wabs*zwabs*pow);
model APL+Gauss is an absorbed power law plus Gaussian absorption lines (XSPEC model wabs*zwabs(pow+zgauss)); APL+Edge is an absorbed power law plus absorption edges 
(XSPEC model wabs*zwabs*zedge*pow); model WARM(APL) is an absorbed power law modified by an intrinsic ionized absorbers (XSPEC model wabs*zwabs*warmabs*pow).
All the energies are rest frame. Errors are given at 90\% confidence level ($\Delta C=2.706$ for one parameter of interest, Avni 1976).}
\begin{center}
\begin{tabular}{cccccc}
\hline\hline\\
\multicolumn{1}{c} {Model} & 
\multicolumn{1}{c} {Parameter} &
\multicolumn{1}{c} {Cha$_{2000}$}  &        
\multicolumn{2}{c} {Cha$_{2007}$} &        
\multicolumn{1}{c} {XMM}      \\
\hline\\ 
1 (APL)  & $\Gamma$                             &   $1.98_{-0.16}^{+0.19}$        & \multicolumn{2}{c}{$2.39_{-0.22}^{+0.22}$}      & $2.06_{-0.13}^{+0.17}$   \\
         & \nh ($\times10^{22}$ cm$^{-2}$)      &    $<1.20$                      & \multicolumn{2}{c}{$4.43_{-1.15}^{+1.20}$}       &  $<0.64$                 \\
         & C-stat/d.o.f.                        &    137.3/170                    & \multicolumn{2}{c}{270.1 / 268 }                 &   493.2/529              \\
\hline\\ 
2 (APL+Gauss) & $\Gamma$                            & $1.80_{-0.15}^{+0.18}$          &  \multicolumn{2}{c}{$2.20_{-0.23}^{+0.22}$}      & $2.04_{-0.15}^{+0.16}$   \\
         &  \nh ($\times10^{22}$ cm$^{-2}$)     &     $<0.89$                     &  \multicolumn{2}{c}{$2.95_{-1.28}^{+1.09}$}      & $<0.41$                   \\
\cline{4-5}
         &  $E_{line}$ (keV)                    & $10.26_{-0.75}^{+0.72}$         &$8.86_{-0.43}^{+0.39}$  &$12.50_{-0.74}^{+0.64}$   &$8.05_{-0.22}^{+0.33}$     \\
         & $\sigma$ (keV)                       &$1.59_{-0.35}^{+0.77}$           & $<0.95$                &$1.34_{-0.61}^{+0.98}$    & $<0.42$                    \\
         &   EW (keV)                           &     $-0.83_{-0.26}^{+0.30}$     & $-0.14_{-0.12}^{+0.09}$&$-0.51_{0.29}^{+0.23}$    &   $-0.13_{-0.04}^{+0.07}$ \\
         & C-stat/d.o.f.                        &    117.0/167                    &   247.0                & 262                      &  478.9/525                \\
\hline\\                                                                                                                           
3 (APL+Edge) &   $\Gamma$                          &  $1.71_{-0.19}^{+0.19}$          &\multicolumn{2}{c}{$1.98_{-0.24}^{+0.25}$}        & $1.90_{-0.16}^{+0.17}$   \\
         & \nh ($\times10^{22}$ cm$^{-2}$)      &  $<0.74$                        &\multicolumn{2}{c}{$2.07_{-1.15}^{+1.24}$}        &$<0.35$                   \\
\cline{4-5}
         &   $E_{edge}$ (keV)                   &  $8.95_{-0.28}^{+0.32}$         &$8.14_{-0.68}^{+0.52}$ &$11.20_{-0.42}^{+0.87}$    & $7.63_{-0.22}^{+0.32}$   \\
         &     $\tau$                           &  $1.85_{-0.83}^{+0.96}$         & $0.60_{-0.33}^{+0.27}$&$0.82_{-0.43}^{+0.49}$    & $1.16_{-0.67}^{+0.94}$   \\
         & C-stat/d.o.f.                        &      120.2/168                  &  246.00&264                                       &  482.8/528             \\
\hline\\                                                                                                                           
4 (WARM(APL))  & $\Gamma$                          & $1.89_{-0.15}^{+0.19}$          &\multicolumn{2}{c}{$2.21_{-0.27}^{+0.32}$}          &$2.35_{-0.19}^{+0.31}$    \\              
          &  \nh ($\times10^{22}$ cm$^{-2}$)    &     $<1.03$                     &\multicolumn{2}{c}{$< 3.63$}                       & $<0.95$                   \\
\cline{4-5}
          &   z                                 &   $1.50\pm0.15$                 & $1.87\pm0.14$ &$0.94\pm0.07$                     & $2.15\pm0.10$             \\
          &   \nh ($\times10^{22}$ cm$^{-2}$)   &      $>55.8$                    &$33.0_{-23.1}^{+79.5}$&$29.3_{-18.6}^{+63.7}$     &$>28.7$                    \\
          & log($\xi$) (erg cm s$^{-1}$)        &   $3.85_{-0.34}^{+1.13}$        & $>3.24$&$>3.25$                                  & $>3.38$                   \\
          & C-stat/d.o.f.                       &   118.3/167                     &    247.6&262                                     &  482.7/527                 \\
\hline      
\hline                              
\end{tabular}
\end{center} 
\end{table*}

\subsection{Absorption edge}

The third model we assumed to fit the spectral features of HS 1700+6416 was Model 1 modified by an absorption edge.
The rest frame edge energy is $E_{edge}=8.95_{-0.28}^{+0.32}$ keV and the maximum absorption depth is $\tau=1.85_{-0.83}^{+0.96}$  for the \chandra\ spectrum of 2000.
For the stacked \chandra\ spectrum, the two edges have $E_{edge1}=8.14_{-0.68}^{+0.52}$ and $E_{edge2}=11.20_{-0.42}^{+0.87}$ keV,
and  $\tau1=0.60_{-0.33}^{+0.27}$ and $\tau2=0.82_{-0.43}^{+0.49}$ respectively.
Finally, the edge parameters for the \xmm\ pn and MOS spectra are  $E_{edge}=7.63_{-0.22}^{+0.32}$ keV and $\tau=1.16_{-0.67}^{+0.94}$.
Given the different shape of the absorption edge, with respect to the Gaussian profile, 
the energies associated to the absorption edges are typically lower than those found for a Gaussian line profile.
The rest frame energies were always consistent with K shell ionization thresholds of mildly/highly ionized gas (Fe XVI-FeXXVI) 
with no blueshifted velocity (Hasinger et al. 2002), except for the highest energy edge of the stacked \chandra\ spectrum, 
at $E_{edge2}\sim11.2$ keV, that require an outflowing velocity of at least v/c=0.18
assuming the recombination energy of hydrogen-like iron, Fe XXVI of 9.28 keV.
However, we don't expect to observe single edges in these kind of source,
but rather complex absorption structures, due to the chaotic nature of the gas, 
that should show a range of ionization parameters and velocities (Ebrero et al. 2011).

\subsection{Warm absorber}

As a further step we used the more physically motivated warm absorber model from XSTAR (Kallman \& Bautista 2001), to modify the primary power law. 
The program computes the spectrum transmitted through a spherical gas shell 
of constant density, thickness $\Delta$R, at a distance R $>>$ $\Delta$R from the
ionizing source. 
In this analysis we used the program implementation that allows to include the XSTAR output directly within
Xspec as a table/model. 
The input parameters of XSTAR are the gas temperature, density, luminosity and spectral shape of the ionizing continuum, turbulence velocity and element abundances,
while the output Xspec model, superimposed to the absorbed power-law of Model 1, has 3 parameters that can be fitted to the spectral data: 
the ionization parameter $\xi$, the absorber column density \nh,  and the absorber redshift (Model 4 in tab. 2).

One important input parameter of the warm absorber model is the turbulence velocity of the gas v$_{turb}$, that controls the broadening 
of the absorption features. 
To reproduce the huge width of the absorption feature in the \chandra\ spectrum of 2000, 
we had to assume a large turbulence velocity, v$_{turb}=3\times10^4$ km s$^{-1}$.
This value is needed to mimic the observed spectral feature but, as said before, it can be interpreted as 
the result of different (contiguous?) gas shells moving with a large range of velocities, or, 
however, as the result of the probably complex dynamics of the wind.

To reproduce the strong feature around the Fe line energies, without producing any detectable absorption at lower energies
the absorber must have high column densities (N$_H>5.5\times10^{23}$ cm$^{-2}$), 
and very high ionization parameter ($log\xi=3.85_{-0.34}^{+1.13}$ erg cm s$^{-1}$),
such that almost all the Fe atoms are in the form of FeXXV-FeXXVI ions (Kallman et al. 2004),
and most of lower-Z elements are completely ionized.
In this model, the absorption feature is reproduced by a blend of the Fe XXV and FeXXVI K$\alpha$
lines. Using the rest frame energies of these lines, i.e. 6.70 and 6.97 keV respectively, 
the redshift of the absorber can be translated into a range of outflowing velocity 
comparable with that obtained from model 2 (v=0.32-0.45c).
The improvement of the fit obtained with this model ($\Delta C\sim19$) is similar to that obtained with the Gaussian absorption line 
for the same number of free parameters.

The same model was applied to the stacked \chandra\ spectrum. The data require two highly ionized absorbers, with similar column densities
($\sim3\times10^{23}$ cm$^{-2}$) and ionization parameters ($Log(\xi)>3.2$  erg cm s$^{-1}$) but outflowing at different velocities,
v1=0.19-0.32c  and v2=0.47-0.58c, to reproduce the features at 2.4 and 3.3 keV, respectively.
In this case, the turbulence velocity required to reproduce the width of the lines is slightly lower, $5\times10^3$ and $10\times10^3$ km s$^{-1}$ respectively.
The goodness of the fit is again comparable to that obtained for the Gaussian absorption line model.
When applied to the XMM spectrum, the model yields a column density \nh$>3\times10^{23}$ cm$^{-2}$, Log$\xi>3.4$ and v=0.12-0.20c,
with an improvement of the fit of $\Delta C/ \Delta \nu= 10.5/3$.

\subsection{Physical parameters of the gas}

Using the values of column density, ionization parameter and outflowing velocities derived from the XSTAR model for the different spectra, together with the 
estimated properties of the source, such as BH mass and X-ray (ionizing) and bolometric luminosity,
we can give a rough estimate of the location, mass outflow rate and kinetic energy associated with the wind.
We adopted the set of equations collected in Tombesi et al. (2012), and based on simple assumptions,
like the definition of ionization parameter and escape velocity for the minimum and maximum distance from the 
central source.
 
The minimum radius, computed as the radius at which the outflowing velocity of the wind equals the escape velocity,
varies in absolute values from the range $2.5-50\times10^{16}$ cm, assuming the very large ($2.5\times10^{10}$ M$_{\odot}$) BH mass estimate available from CIV,
to $0.4-8.4\times10^{16}$ cm, using the lower limit for the BH mass ($3.9\times10^{9}$ M$_{\odot}$), estimated assuming an Eddington rate accretion.
In both cases (and independently of the BH mass) the very high outflowing velocities measured, imply an origin of the wind extremely close to the BH,
i.e.  in the range 3-70 $r_s$.
The maximum radius can be inferred from the definition of ionization parameter, 
as the maximum distance at which the ionizing luminosity can produce a ionization $\xi$ in a gas of column density \nh:
$r_{max}=L_{ion}/\xi N_{H}$.
The ionizing luminosity taken into account by XSTAR when computing the properties of the ionized gas 
is defined as the unabsorbed luminosity of the source in the range 
13.6 eV-13.6 keV.
L$_{ion}$ is in the range $0.8-6.1\times10^{46}$ ergs s$^{-1}$, and the associated
radius is comprised between $0.2-8.9\times10^{19}$ cm, i.e. $2.6\times10^2-1\times10^4 r_s$ for a BH mass of $2.5\times10^{10}$ M$_{\odot}$,
and $1.6\times10^3-7.4\times10^4  r_s$ for a  BH mass of $3.9\times10^{9}$ M$_{\odot}$.
The values derived for the Chandra data of 2007 and the XMM data are upper limits, due to the lower limits in ionization parameter.
The results obtained for $r_{max}$ are consistent with what found in Tombesi et al. (2012), for a sample of lower luminosity, local Seyferts,
while the higher outflowing velocities observed in HS 1700+6416, with respect to their results, translates into lower values
of $r_{min}$, i.e. the wind can originate closer to the BH.

To compute the mass outflow rate we used the expression derived by Krongold et al. 2007 for a biconical wind, also used in Tombesi et al. 2012,
 $\dot{M}_{out}=0.8 \pi m_p N_H v_{out} r f(\delta, \phi)$. $f(\delta, \phi)$ depends on the relative orientation of the disk, the wind and the line of sight,
but for reasonable angles can be considered of the order of unity.
The formula assumes a conical geometry with thickness constant and negligible with respect to the distance of the wind from the central source.
The minimum mass outflow rates, associated with the minimum radius estimated above,
are $\dot{M}_{out}(r_{min})=4-6 M_{\odot}/yr$, and the associated kinetic energies are in the range 
$\dot{E}_{K}(r_{min})=0.3-27\times10^{46}$ erg s$^{-1}$. 
These values translate into a minimum $\dot{M}_{out}/\dot{M}_{acc}=0.03-0.13$
and a minimum $\dot{E}_{K}/L_{bol}=0.01-0.18$,
computed assuming for each observation, the $L_{bol}$
derived from the observed X-ray luminosity, an X-ray bolometric correction of 50, 
an efficiency $\eta=0.05$, and constant density and velocity.
Unfortunately, the maximum mass outflow rates remain unconstrained, due to the combination of 
lower limits in \nh\ and $\xi$.
However, such high minimum values are already very large, telling us that, if persistent and corresponding to the simple geometry assumed here, 
the massive, fast outflowing wind observed in  
HS 1700+6416 carries a huge amount of energy, and is perfectly capable of providing the mechanical power required to produce a significant feedback in the environment,
estimated to be of the order of $\sim5\%$ of the bolometric luminosity  (Di Matteo et al. 2005; Ostriker et al. 2010),
or even a order of magnitude less in the case of multistage feedback (Hopkins \& Elvis 2010).
We also underline that the highest values of kinetic energy and mass outflow rate come from the 
feature with the most secure detection, i.e. from the \chandra\ spectrum of 2000, making our results more robust.
Understanding the physics of these winds is necessary to obtain tighter constraints on their properties and on their impact on the surrounding environment.

\section{Conclusions}

We performed a uniform and comprehensive analysis of all the \chandra\ and \xmm\ observations, taken between 2000 and 2007,
and determined the X-ray properties of the high redshift NAL-QSO HS 1700+6416.
The source was found to be constant  in flux and spectral parameters on timescales of tens of ks, i.e. the duration of the longest
observation available.
Furthermore, the X-ray spectral parameters were constant during the 
8 contiguous observations of 2007, performed 
in a time interval of 9 days.
This allowed the stacking of the 8 spectra taken in 2007, resulting in a larger signal-to-noise spectrum for that period.
The BH mass estimate available from the SDSS spectrum is extremely high (M$_{BH}$=$2.5\times10^{10}$ $M_{\odot}$)
and translates into a variability time scale of $\sim10$ days in the observer frame.
Using the bolometric luminosity and Eddington limit accretion, we derive  
a lower limit for the BH mass of $3.9\times10^{9}$ $M_{\odot}$, implying a variability time scale of $\sim1.6$ days.

On longer timescales, the source is found to be variable on timescales of years, both in the observed 2-10 keV flux, that varies by a factor of 3 
between  2002 and 2007, and in the amount of neutral absorption, that is negligible in 2000 and 2002,
and becomes of the order of $5\times10^{22}$ cm$^{-2}$ in 2007.

The most remarkable feature however is the detection of a strong, broad and variable absorption feature, 
observed at energies higher than that of the neutral Fe K$\alpha$ emission line and up to $\sim12$ keV.
The strong, broad absorption feature is clearly detected, at high significance ($>4\sigma$), in the spectrum extracted from the longest
observation available, at an energy of $\sim10.2$ keV. 
In the stacked spectrum of 2007 a similar feature is detected ($>3\sigma$) at higher energies ($\sim12.5$ keV).
Furthermore, a hint of a feature at $\sim8-9$ keV is observed, albeit at lower significance ($\sim2\sigma$), both in 
the stacked \chandra\ spectrum of 2007 and in the short ($\sim10$ ks) \xmm\ spectrum of 2002.
Finally, a clear variability is detected in the energy of the line, at least between the two strongest features 
observed at 10.2 and 12.5 keV (fig. 6).

In addition to the phenomenological model (a power-law modified by a Gaussian absorption lines) used
to determine the significance of the features,
we also used a model including absorption edges instead of Gaussian lines.
The energies obtained are consistent with 
ionization thresholds of mildly/highly ionized gas (Fe XVI-FeXXVI) 
with no outflowing velocity, except for the  highest energy edge of the stacked \chandra\ spectrum, 
for which a velocity v$\sim0.2$c is required.

Finally, we adopted a more physical approach, modeling the absorption with the XSTAR photo-ionization code.
The fit with XSTAR requires, for all the spectra, a dense (\nh$>3\times10^{23}$ cm$^{-2}$) highly ionized
(log$\xi>3.2$ erg cm s$^{-1}$) and highly turbulent (v$_{turb}$=$0.01-0.1c$) absorber, such that a blend
of Fe XXV and Fe XXVI  K$\alpha$ absorption lines reproduces the spectral feature.
The large turbulence velocity is required to reproduce the extreme width of the lines, however this broadening can be interpreted
as the result of the superimposition of different gas shells, with different velocities, rather than 
as an internal turbulence.
The inferred outflowing velocities are in the range v=0.12-0.59c

All these pieces of information make the NAL-QSO HS 1700+6416 
the best candidate to be the first non-lensed quasar showing 
X-ray BALs.
The measured properties of this source in velocity range, line width and inferred physical parameters of the obscuring gas,
are comparable only with those reported for the X-ray BALs APM 08279+5255 and PG 1115+080 (Chartas et al. 2002, 2003).
In particular APM 08279+5255 has BH mass ($M_{BH}\sim2\times10^{10} M_{\odot}$) and Eddington ratio ($L_{Bol}/L_{Edd}\sim0.3$)
very similar to those of HS 1700+6416. 
Also similar is the large difference between the X-ray and UV absorption systems outflowing velocities,
(i.e. a ratio of the order of ~10-15) possibly implying that
UV and X-ray BAL are produced by different absorbers at different distances from the central BH.

With the parameters obtained from the fit comprising the ionized absorber, and assumptions on the geometry of
the wind and the efficiency of the accretion,
we were able to derive a rough estimate of the location of the wind,
that spans the range $3-10^4$ $r_s$, 
and of the minimum mass and kinetic energy content of the wind itself, 
that varies from few \% up to $\sim13$ and $\sim18\%$ of the mass accretion rate and Bolometric luminosity respectively.
Even if only approximated, these results, together with others, emerging from the analysis of local sources 
as well as from large blind search surveys,
are starting to quantitatively probe that high ionized, massive, fast outflowing disk winds can provide the mechanical 
energy needed to produce the feedback that might be responsible for the well known relation between galaxies properties
and BH masses.

\begin{acknowledgements}

We thank the referee for useful comments that improved the paper.
The author thanks P. Grandi, E. Torresi and G. G. C. Palumbo 
for useful discussions.
Partial support from the Italian
Space Agency (contracts ASI/INAF/I/009/10/0) is acknowledged. 
This research has
made use of the NASA/IPAC Extragalactic Database (NED) which
is operated by the Jet Propulsion Laboratory, California Institute
of Technology, under contract with the National Aeronautics and
Space Administration, and of data obtained from the Chandra
Data Archive and software provided by the Chandra X-ray Center
(CXC). Also based on observations obtained with XMM-Newton, an ESA science mission
with instruments and contributions directly funded by
ESA Member States and NASA.

\end{acknowledgements}

\begin{appendix}

\end{appendix}

\end{document}